\documentclass[twocolumn,prd,showpacs,superscriptaddress,groupedaddress,preprintnumbers,nofootinbib]{revtex4-1}
\pdfoutput=1
\usepackage{graphicx,amsfonts,amsmath,amssymb,epsfig,color, mathrsfs,latexsym,url,wasysym,float}
\newcommand{\be}{\begin{equation}}
\newcommand{\ee}{\end{equation}}

\newcommand{\ba}{\begin{eqnarray}}
\newcommand{\ea}{\end{eqnarray}}
\newcommand{\mpl}{m_{\rm Pl}}

\begin{document}

\title{Electroweak Vacuum Stability and the Higgs Field Relaxation\\ via Gravitational Effects}

\author{Mahdi Torabian$^*$}
\address{Department of Physics, Sharif University of Technology, Azadi Ave, 1458889694, Tehran, Iran}

\begin{abstract}
The measured values of the Standard Model (SM) parameters favors a shallow metastable electroweak (EW) vacuum surrounded by a deep global AdS or a runaway Minkowski minimum. Furthermore, fine-tuning is the only explanation for the Higgs relaxing in its present local minimum. In this paper, assuming no new physics beyond the SM, we study the universal effect of gravity on the Higgs dynamics in the early universe. A generic two-parameter model is considered in which the Higgs is non-minimally coupled to a higher-curvature theory of gravity. The coupling between the Higgs field and the Weyl field in the Einstein frame has genuine predictions. In a broad region in the parameter space, the effective Higgs mass is large and it initially takes over through fast oscillations. This epoch is followed by the Weyl field slowly rolling a plateau-like potential. This framework generically predicts that the Higgs self-coupling in the EW vacuum is enhanced, compared to the SM predictions, through couplings to the gravity sector. Moreover, when the Higgs is settled in the EW vacuum, all other scalar flat directions would be lifted via gravitational effects mediated by the Weyl field. 
\end{abstract}

\preprint{SUT/Physics-nnn}
\maketitle

\subsection*{I. Introduction} 
The great achievement of the LHC has been the discovery of the Standard Model (SM) Higgs boson \cite{Aad:2012tfa,Chatrchyan:2012ufa,Aad:2015zhl} with mass $m_{h} = 125.35\pm0.15$  GeV \cite{CMS:2019drq}. 
The scalar sector of the SM is completed and all parameters are determined. In particular, the Higgs self-coupling parameter is deduced at the electroweak scale to be around $\lambda(m_{\rm EW})  \approx 0.13$. 
This is the only parameter of the SM which is not multiplicatively renormalized. 
With the central value of top quark mass $m_t = 173.2\pm0.9$ GeV \cite{ATLAS:2014wva}, the beta-function $\beta_\lambda$ (at low/intermediate scales) is dominated by the top Yukawa coupling and thus it is negative. 
The SM, as a renormalizable theory, can in principle be applied in an arbitrary high energy and make predictions. If fact, the LHC has found no trace of new physics and no significant deviation of the SM predictions are observed. Within the SM, the effective Higgs potential can be computed at desired loop orders. The self-coupling parameter is monotonically decreasing and it vanishes at an intermediate energy around $10^{11}$ GeV and subsequently turns negative (see Fig. 1) \cite{Degrassi:2012ry,Buttazzo:2013uya,Bednyakov:2015sca}. 
\begin{figure}[b!]
\begin{center}\includegraphics[scale=.32]{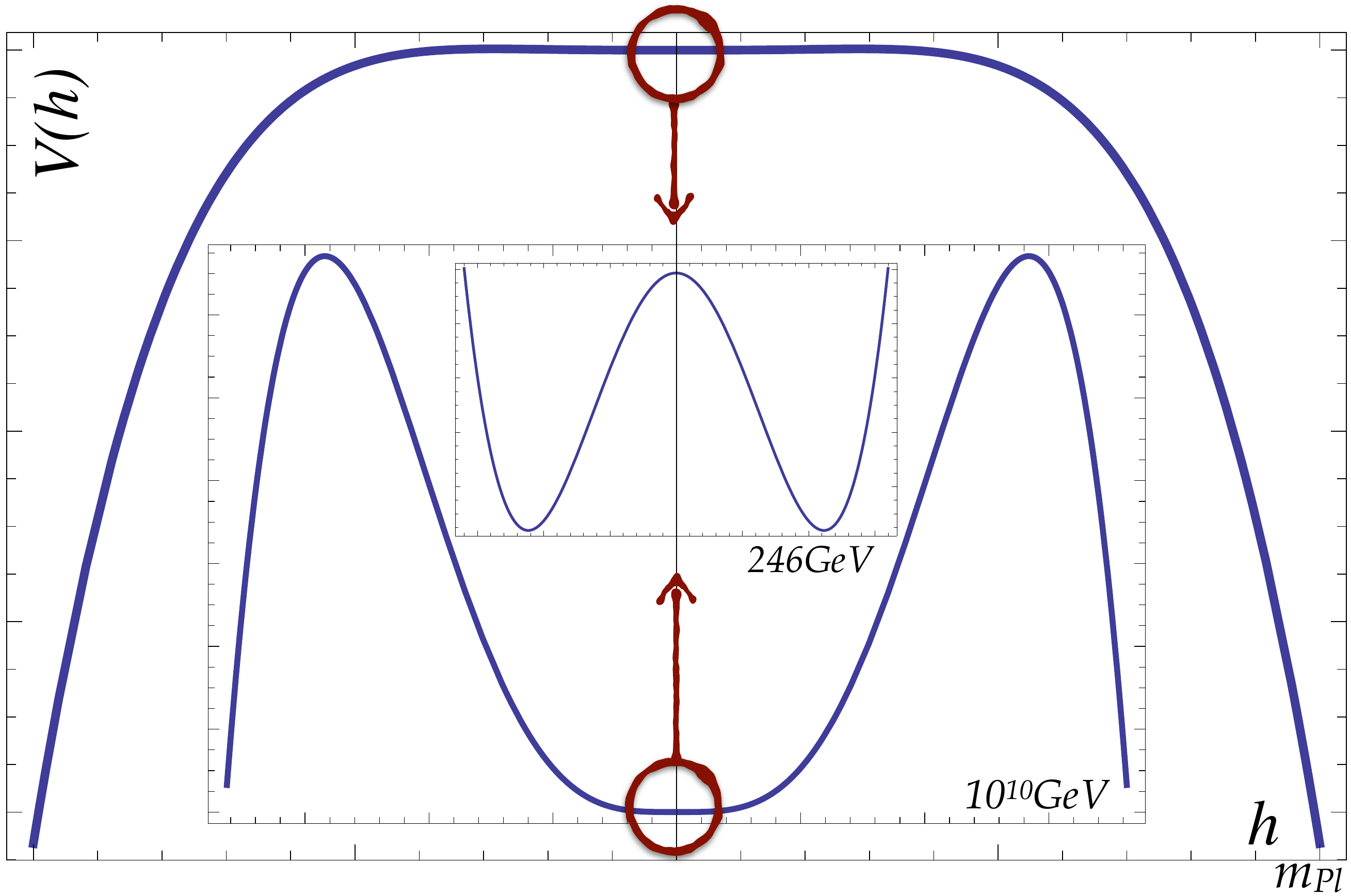}\end{center}
\caption{The shape of the Higgs potential in the SM up to the Planck scale. Inner panels show zoomed-in regions}\end{figure}

At higher scales the gauge interactions take over and make the beta function positive. Then, the quartic coupling is increasing as it develops a new minimum which will be the global one. 
The location of the global minimum is located at tens of $\mpl$ and is sensitive to Planck suppressed operators. It is reasonable to abandon the naive extrapolation to an arbitrary high energy and limit the running up to the Planck scale. Thus, the potential would basically be seen ill as it is unbounded from below. The electroweak vacuum is a local minimum and the barrier separating it from the deep well is extremely small. If one computes the tunneling rate between the vacua and ignores Planck suppressed interactions, one finds that life-time of the present-day electroweak vacuum is greater than the age of the Universe and thus the vacuum is {\it metastable} \cite{Sher:1988mj}. However, if one includes higher dimensional operators, the the lifetime would be much shorter \cite{Branchina:2013jra,Burda:2015isa}. Consequently, due to a huge negative cosmological constant, it leads to a catastrophic gravitational collapse.

Moreover, the Higgs potential raises issues in connection to the early universe cosmology.
In order to end up in the present-day electroweak vacuum and prevent the Higgs from rolling down to global AdS minimum, a fine-tuning at level of {\it one} part in {\it a hundred million} in the Higgs value is needed \cite{Lebedev:2012sy,Bars:2013vba}. Moreover, if that initial condition is prepared, the Higgs will not stick to that in the presence of Hubble-size quantum fluctuations during a high scale inflation.

New physics beyond the SM, including new particles and/or new interactions, could possibly change this picture and stabilize the Higgs potential. However, excellent agreement of the SM predictions with the experimental results puts tight constraints on new physics as it must have marginal effect on the electroweak fit. Moreover, generically new physics would inevitably introduce a naturalness problem to the scalar sector. 

The scalar fields generically have unsuppressed couplings to other fields and provides sizable portals to the different sectors (the very same feature that causes instability and unnaturalness of the Higgs). In particular the scalar and tensor backgrounds have sizable interactions. In this paper, irrespective of presence or absence of new physics beyond the SM, we study the ubiquitous effect of gravity on the Higgs dynamics in the early universe. We consider a well-motivated framework in which the Higgs field is non-minimally coupled to a higher-curvature theory of gravity (see \cite{Bamba:2015uxa,Salvio:2015kka,Calmet:2016fsr,Ema:2017rqn,Ema:2017loe,Ema:2020evi,Ema:2019fdd,Ema:2020zvg,Herranen:2015ima,Markkanen:2018bfx,Markkanen:2018pdo} for recent studies of different aspects of these couplings). There are two gravitational free parameters in this framework and different dynamics can be found in different regions of the parameter space. Through a conformal transformation, we can move to the Einstein frame which has direct contact to observables. In these coordinates, there are genuine couplings between the Higgs field and the emergent Weyl field. Both fields have a plateau-like potential. It has been know that the Higgs field itself can play the role of inflaton \cite{Bezrukov:2007ep}
. In this framework, it is definitely possible is a corner in the parameter space. The Higgs inflation at tree-level is in perfect agreement with the observation of the CMB spectrum as it accommodates the spectral index of scalar power spectrum and the tensor-to-scalar ratio. However, there are debates that quantum loop effects might jeopardize predictions and make the scenario complicated \cite{Barbon:2009ya,Barvinsky:2009ii,Bezrukov:2010jz}. These effects include the above mentioned instability of the potential and the violation of perturbative unitarity close to the inflationary scale \cite{Burgess:2009ea,Burgess:2010zq,Hertzberg:2010dc}
.  

These complications can be avoided in other regions of the parameter space which is the aim of this paper. The effective curvature of the Higgs potential for a broad range of parameters and large field values is large. Initially in the early universe, the Higgs field dominates the dynamics as it coherently oscillates about its minimum. The universe is matter dominated and the energy in the Higgs field is drifted away by cosmic expansion. The Higgs-Weyl interactions alleviate the instability problem and eventually the Higgs field is settled close to its present-day electroweak values. Finally, the Weyl (inflaton) field takes over the dynamics and its plateau-like potential derives cosmic inflation \cite{Starobinsky:1980te} which is in great agreement with inflationary observables in recent {\sl Planck} results \cite{Akrami:2018odb}. Moreover in this framework, the structure of the electroweak vacuum is modified gravitationally compared to the SM. In particular the Higgs self-coupling parameter receives contributions from the gravity sector. In general we observe that in this setup, through omnipresent Weyl-scalars interactions, all scalar fields develop non-flat potentials with masses and may also receive non-zero VEV's.

The structure of the paper is as follows. 
In the next section we introduce the model via its classical action. Then we study the stability condition by analyzing the scalar potential in the Einstein frame. Next we numerically solve the equations of motion. Then we study physics around the electroweak vacuum and compute the Higgs sector parameters. Next we show that in this framework all moduli are lifted and there is no flat directions. Finally, we conclude in the last section.

\subsection*{II. The Action}
The dynamics of the Higgs field which is non-minimally coupled to a higher-curvature theory of gravity is given by the following action parametrized in the Jordan frame
\ba S=\int {\rm d}^4x(-g_J)^{1/2}\Big[ &&{\textstyle\frac{1}{2}}\big(m^2+\xi\phi^2\big)R_J+{\textstyle\frac{1}{4}}\alpha R_J^2 \cr &&- {\textstyle\frac{1}{2}} g_J^{\mu\nu}\partial_\mu\phi\partial_\nu\phi -V_J(\phi)\Big] ,\ea
where $\phi^2 = 2 H^\dagger H$. The action includes all the operators up to dimension four which respects gauge symmetries. Thus, they must be included in a consistent quantum theory as they are needed based on perturbative renormalization theory \cite{Starobinsky:1980te,Callan:1970ze}. In this framework, the parameters $\xi$ and $\alpha$ define a two-parameter family of models. Needless to say, physics is different in different regions of the parameter space. The non-minimal scalar-gravity coupling is studied in variety of models especially connected to cosmic inflation. Moreover, the term quadratic in the Ricci scalar is the simplest generalization to General Relativity. Although it is a higher-derivative theory of gravity, it is free from Ostrogradski classical instability or the presence of spin-2 ghost (and also spin-0 ghost for positive $\alpha$) in the spectrum \cite{Stelle:1976gc}.  

To make direct contact with observables, we can move to the Einstein frame through a conformal transformation of the metric
\be g_{\mu\nu}^E = \mpl^{-2}(m^2+\xi\phi^2 + \alpha R) g_{\mu\nu} \equiv e^{\sqrt{{\textstyle\frac{2}{3}}}\mpl^{-1}\chi} g_{\mu\nu}.\ee
Then, the action is
\ba\label{Einstein-frame-action} S_E = \int {\rm d}^4x(-g_E)^{1/2}{\textstyle\frac{1}{2}}\Big[\mpl^2R_E - g_E^{\mu\nu}\partial_\mu\chi\partial_\nu\chi\!\!\!\!\!&& \cr -e^{-\sqrt{{\textstyle\frac{2}{3}}}\mpl^{-1}\chi}g_E^{\mu\nu}\partial_\mu\phi\partial_\nu\phi  -2&&V_E(\phi,\chi)\Big].\ \ \ \ea
The Ricci-squared term introduces a new propagating scalar field, {\it a.k.a.} Weyl scalar. In fact, the higher derivative term make a spin-0 degree of freedom propagating which is not ghost-like. It is not seen in the Jordan frame and is transparent in the Einstein frame. Note that the Weyl scalar has a canonical kinetic term while the Higgs field is non-canonical. 
In fact, the Weyl and the Higgs fields interact via the derivative terms besides the scalar potential.
The scalar potential in the Einstein frame reads as
\ba\label{potential} V_E(\phi,\chi)&=&e^{-2\tilde\chi}\Big[{\textstyle\frac{1}{4}}\alpha^{-1}\big(\mpl^2e^{-\tilde\chi}-m^2-\xi\phi^2)^2+V_J(\phi)\Big]\cr  &=&e^{-2\tilde\chi}\Big[{\textstyle\frac{1}{4}}\alpha^{-1}\big(\mpl^2 e^{\tilde\chi}-m^2\big)^2 + {\textstyle\frac{1}{4}}(\lambda+\xi^2\alpha^{-1})\phi^4\cr &&\qquad\ +{\textstyle\frac{1}{2}}\big(\mu^2-\big(\mpl^2 e^{\tilde\chi}-m^2\big)\xi\alpha^{-1}\big)\phi^2\Big],\ea
where $\tilde\chi=\sqrt{{\textstyle\frac{2}{3}}}\mpl^{-1}\chi$. We also introduced the conventional Higgs potential (with a mass parameter $\mu^2<0$)
\be V_J(\phi)=\frac{1}{2}\mu^2\phi^2+\frac{1}{4}\lambda\phi^4.\ee
As can be seen from the potential \eqref{potential}, the Higgs mass parameter and the quartic coupling in the Einstein frame received contributions from the gravitational sector. 

There is an upper bound on the value of $\alpha$ around $\alpha\lesssim 10^{61}$ from gravitational experiments measuring Yukawa correction to the Newtonian potential \cite{Hoyle:2004cw,Calmet:2008tn}. For greater values, the mass of the Weyl field is less than the present Hubble rate around $10^{-33}$ eV.  Moreover, the parameter $\xi$ basically normalizes the 4-dimensional Planck mass in the Einstein frame. An upper limit exists only when it is positive  $\xi \lesssim (\mpl/v)^2\sim 10^{32}$.  A much tighter bound can be put via collider physics. As argues above, we need to rescale the Higgs field to make its kinetic term canonical. Around the electroweak vacuum we find that
\be \varphi \equiv e^{-\tilde\chi_0/2}\phi \approx (1+\xi v^2/\mpl^2)\phi.\ee
Therefore, the Higgs coupling to the SM particles is modified. This modification has an observable effect at colliders by suppressing or enhancing the decay modes of the Higgs particle. 
The combined analysis of the ATLAS and CMS excludes $|\xi|\gtrsim 10^{15}$ at $95\%$ C.L.  \cite{Atkins:2012yn}.  Thus, we find a large (gravitational) parameter space for $1\lesssim\alpha\lesssim 10^{61}$ and $|\xi|\lesssim 10^{15}$ and different parameters, different dynamics can be obtained. 
As argued in introduction, we are interested in values through which the cosmic inflation is driven by the plateau-like potential of the Weyl field.  The {\sl Planck} results on the CMB anisotropy $\log(10^{10}A_s)=3.044\pm0.414$ 68\% C.L.  and the primordial gravitational waves $r<0.11$ 95\% C.L. \cite{Akrami:2018odb} constraint the free parameter $\alpha$ as
\be \alpha = (12\pi^2 r A_s)^{-1}\gtrsim 3.4\times 10^7. \ee
The simplest manifestation of the Starobinsky inflation predicts $r\approx2.5\times 10^{-3}$ and therefor $\alpha\approx  10^9$. However, modifications to the model predict larger $r$ and so smaller $\alpha$ works as well (see \cite{Ben-Dayan:2014isa}). As we later see, the electroweak vacuum further constrains the parameter space.  

\subsubsection*{Stability conditions}
As can be seen from the scalar potential \eqref{potential} stability at Planck field values can be obtained for
\be\label{condition-1} \lambda(\mpl)+ \xi(\mpl)^2\alpha(\mpl)^{-1} \geq 0.\ee
Assuming no new physics between the electroweak scale and the Planck scale and applying the central values of the measures SM parameters, the best-fit value for the Higgs self-coupling at the Planck scale is 
\be \lambda(\mpl)\approx -0.0129.\ee
Then, the stability condition \eqref{condition-1} implies that ($\alpha\approx10^9$)
\be |\xi(\mpl)|\gtrsim {\rm few}\times 10^3.\ee
For negative $\xi$ the Higgs potential is convex for any value of the scalar fields. For positive $\xi$,  further condition on initial field values is imposed so that the quadratic Higgs term does not take over the quartic term to  destabilize the potential
\be \tilde\chi_{\rm ini}\lesssim 2 \ln\tilde\phi_{\rm ini} + \ln(\xi/2).\ee
Similarly for negative $\xi$, if the initial Weyl field value satisfies
\be \tilde\chi_{\rm ini} \gtrsim 16.1 + 2 \ln\tilde\phi_0 - \ln(-\xi),\ee
then Higgs quadratic term takes over the quartic term and makes the Higgs potential stable in large field values. It helps to choose smaller value of $|\xi|$. In the rest of the paper, for concreteness, we choose positive values of $\xi$ and study the evolution of the Higgs and the Weyl fields.

\subsection*{III. Dynamics in the Early Universe}
\begin{figure*}[t!]
\begin{center}
\includegraphics[scale=.5]{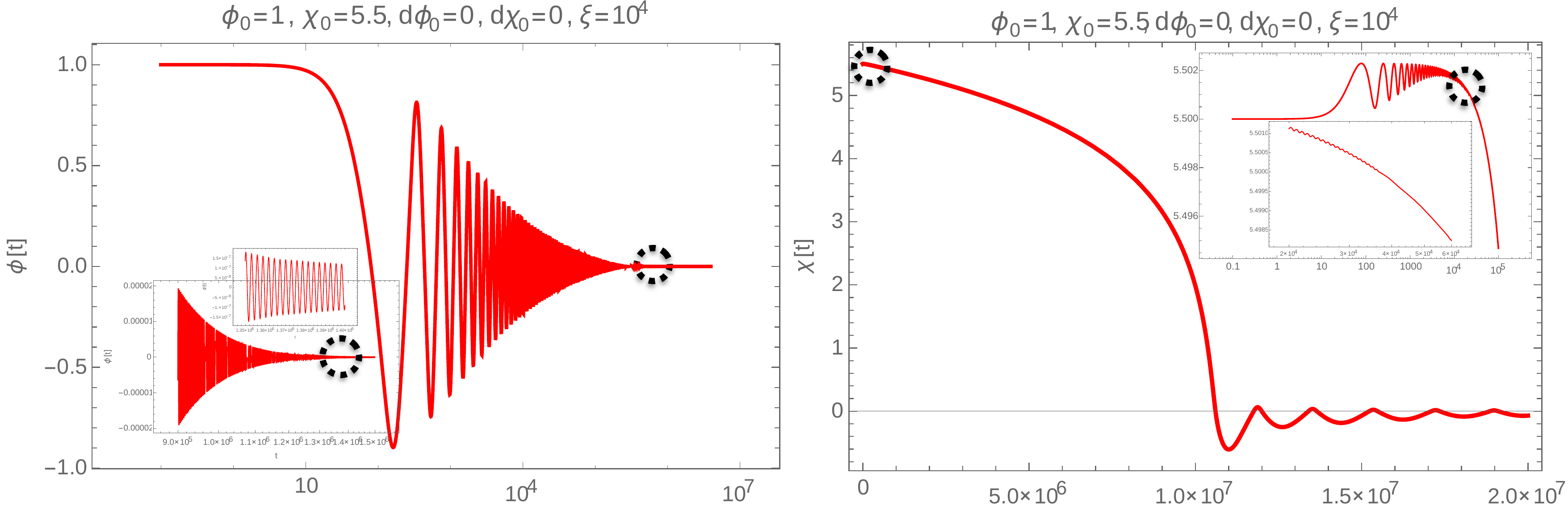}
\end{center}
\caption{Time evolution of the Higgs (left panel) and the Weyl (right panel) fields. The initial conditions are given on the top.}\end{figure*}
Assuming spatial homogeneity, the dynamics of the Higgs field $\phi(t)$, the Weyl field $\chi(t)$ and the scale factor $a(t)$ in the Friedman metric 
\ba {\rm d}s^2 = -{\rm d}t^2 + a(t){\rm d}^2{\bf x},\ea
is governed by the following equations of motion
\ba && \ddot\chi + 3H\dot\chi + \frac{1}{\sqrt6}e^{-\tilde\chi}\dot\phi^2+V^E_{,\chi}=0 ,\\
&& \ddot\phi + 3H\dot\phi - {\textstyle\sqrt{\frac{2}{3}}}\mpl^{-1} \dot\chi\dot\phi +V^E_{,\phi}=0,\\ 
&&3 H^2 \mpl^2 = \frac{1}{2}\dot\chi^2 + \frac{1}{2}e^{-\tilde\chi} \dot\phi^2 + V_E,\\
&&-2 \dot H \mpl^2 = \dot\chi^2 + e^{-\tilde\chi} \dot\phi^2.\ea
In the above equations $V^E_{,\phi}$ and $V^E_{,\chi}$ are field derivative of the scalar potential and $H=\dot a/a$ is the Hubble expansion rate.
These are coupled second-order differential equations that can be solved by numerical methods. The solutions for scalar fields in Planck mass versus Planck time are plotted in figure 2. 
The Higgs field initial value is taken of order one in Planck mass. Its initial velocity could also be chosen order one, however, it is found that it has insignificant qualitative effect on the solutions. On the other hand, the Weyl field initial conditions are chosen such that the universe undergoes at least 60 e-folds of exponential expansion.

The solutions are interpreted as follows. The Higgs and the Weyl fields are initially frozen for tens of Planck time (mini-inflation) until they commence harmonic oscillations about their local minima. The effective Higgs mass is large and the Higgs field oscillates with large amplitudes. The energy density in its coherent oscillations takes over the dynamics of the universe and it is redshifted away by cosmic expansion. The Higgs field values is decreasing and relaxing toward its small values.  It is important to emphasis that by this time the Higgs field amplitude is less than $10^{-8}\mpl$ so it later evolves to the electroweak vacuum. 
When the Hubble rate is around $10^{-4}\mpl$, the Weyl field takes over the energy density by its plateau-like potential and slowly rolls down. The universe enters an epoch of inflation which lasts  around 60 e-folds.  Then, the Weyl field oscillates about its minima and the universe is filled by the Bose condensates of Higgs and Weyl particles. 
After many damped oscillations fields settle down in their minima near the origin. Finally, they decay and reheat the universe.  
The fields have slightly different evolution, although qualitatively the same, depending on the value of the non-minimal coupling parameter as is plotted in figure 3.

\begin{figure*}[t!]
\includegraphics[scale=.5]{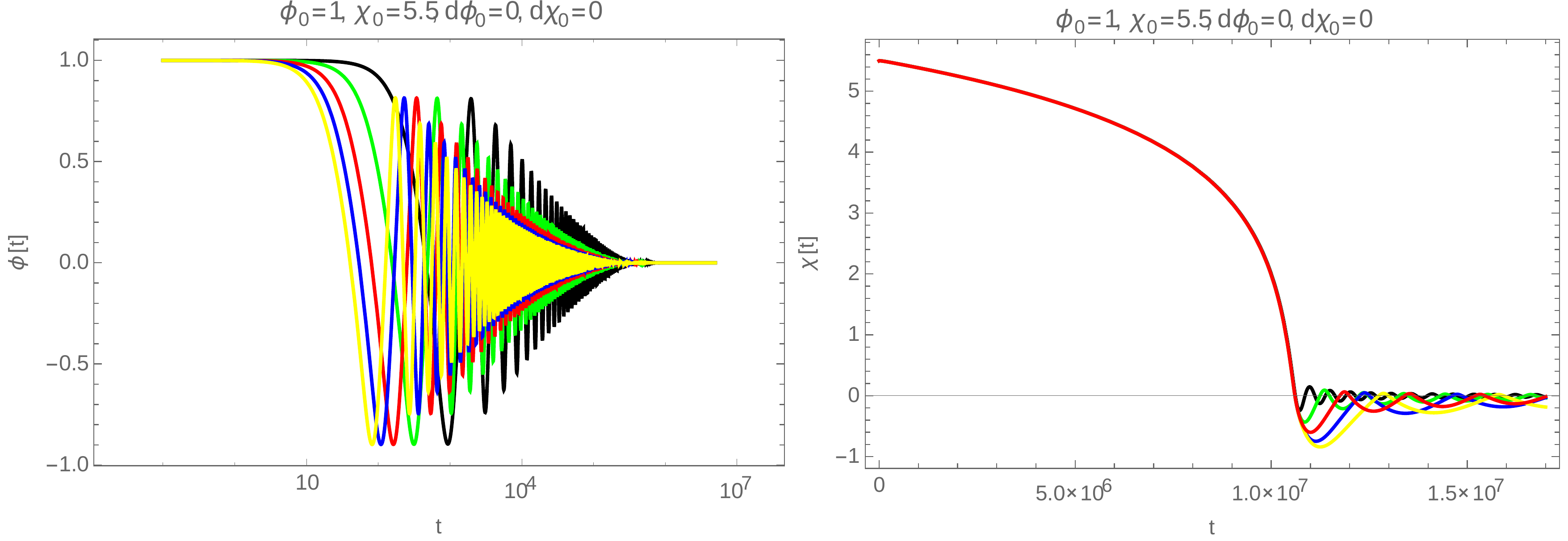}
\caption{solutions for different $\xi$ parameter: $\xi$=1000(black), 5000(green), 10000(red), 15000(blue), 20000(yellow)}\end{figure*}

\subsection*{IV. Predictions for the Electroweak Vacuum}
At late times, the Higgs field is closed enough to the origin of its potential and, through the symmetry breaking mechanism, is settled to its local minimum. Via appropriately choosing the mass parameter, the Higgs field receives a non-zero vacuum expectation value (VEV) and spontaneously breaks the electroweak symmetry. The VEV and the curvature about the minimum are respectively computed as
\ba\label{mphi} v^2\equiv\phi_0^2 &=& \frac{-\mu^2}{\lambda+\xi\frac{-\mu^2}{m^2}}\approx \frac{-\mu^2}{\lambda}\approx(246\ {\rm GeV})^2,\\
m_\phi^2 &=&  -2\mu^2\frac{\lambda+\xi^2\alpha^{-1}}{\lambda+\xi\frac{-\mu^2}{m^2}}\approx-2\mu^2(1+\xi^2\lambda^{-1}\alpha^{-1}).\ \ \ \ \ea
The Higgs VEV in the first equation is fixed by the measured masses of particles in the SM. As $m^2\gg-\mu^2$ the sign of the Higgs non-minimal coupling parameter is not relevant in the symmetry breaking mechanism. Furthermore, the Weyl field receives a non-zero VEV and mass as follows
\ba\label{chi-vev} \chi_0 &=& {\textstyle\sqrt{\frac{2}{3}}}\mpl\ln\Big[\frac{m^2}{\mpl^2}+\frac{-\mu^2}{\mpl^2}\frac{\xi-\alpha\frac{-\mu^2}{m^2}}{\lambda+\xi\frac{-\mu^2}{m^2}}\Big]\cr &\approx& \xi\frac{-\mu^2}{\lambda}\mpl^{-1} \approx 10^{-5}\xi\ {\rm eV},\\
m_\chi^2&=&\frac{\mpl^2}{3\alpha}\Big[\frac{m^2}{\mpl^2}-\frac{\mu^2}{\mpl^2}\frac{\xi}{\lambda-\xi(\mu^2/m^2)}\Big] \cr &-&\mu^2\frac{\mu^2}{\mpl^2}\frac{2-\frac{\lambda}{\lambda-\xi(\mu^2/m^2)}}{3(\lambda-\xi\frac{\mu^2}{m^2})}\approx \frac{\mpl^2}{3\alpha} \sim 10^{13}\ {\rm GeV}.\ \ \ \ea
The interactions between the Higgs and the Weyl fields besides the Higgs receiving a VEV induces a non-zero VEV for the Weyl field. It is sensitive to the sign of the non-minimal coupling parameter. It further implies that the canonical Higgs field is obtained by a field redefinition
\be \phi_c = e^{-\tilde\chi_0/2}\phi.\ee

As a result of Higgs-Weyl interactions, there is also a non-diagonal component in the scalar mass matrix 
\ba m_{\rm mix}^2=V_{,\phi\chi}&=&-{\textstyle\sqrt{\frac{2}{3}}}e^{-{\textstyle\sqrt{\frac{2}{3}}}\mpl^{-1}\chi_0}\xi\alpha^{-1}v\mpl\cr &\approx& -\sqrt{\frac{-2\mu^2}{3\lambda}}\xi\alpha^{-1}\mpl.\ea
Upon diagonalizing the mass matrix by a rotation
\be {\rm diag}(m_{\rm Weyl}^2, m_{\rm Higgs}^2) = R(-\theta)[V''(\phi,\chi)]R(\theta),\ee
we find the following mass eigenvalues
\ba m_{\rm Weyl}^2&\approx&\frac{\mpl^2}{3\alpha} \sim (10^{13}\ {\rm GeV})^2,\\
m_{\rm Higgs}^2 &\approx& -2\mu^2\sim (125\ {\rm GeV})^2.\ea
The former is the inflaton mass which is effectively decoupled from low-scale physics. 
The lighter mass eigenvalue is the mass of the Higgs particle and is compared with its measured value. Then, the Higgs mass parameter and the self- coupling are determined as follows
\ba \mu^2(m_{\rm EW})&\approx& -(89.1\ {\rm GeV})^2,\\ 
\label{lambda}\lambda(m_{\rm EW}) &\approx& 0.13.\ea

The mass-mixing angle is negligible and computed as
\be \tan2\theta\approx -{\textstyle\sqrt 6}\xi v\mpl^{-1}\approx -10^{-16}\xi.\ee
The {\it physical} Higgs field is the mass eigenvector
\be h = \cos\theta e^{-\tilde\chi_0/2}\phi + \sin\theta\, \chi.\ee
The couplings of the Higgs boson to other particles (compared to the SM) are suppressed due to gravitations effetcs. Thus, the Higgs production and decay rates are affected. It has an observable effect at colliders in terms of  the global signal strength
\ba \mu\equiv{\textstyle\frac{\sigma\cdot{\rm Br}}{\sigma_{\rm SM}\cdot{\rm Br}_{\rm SM}}} = e^{-\tilde\chi}\cos^2\theta
\approx 1-6\mpl^{-2}\xi^2v^2.\ea
The combined analysis of the ATLAS and CMS measurements implies $\mu= 1.07\pm0.18$  \cite{Aad:2012tfa,Chatrchyan:2012ufa,Atkins:2012yn}. 
Therefore, this measurement excludes $|\xi|\gtrsim 10^{15}$ at $95\%$ C.L. However, a tighter bound is obtained from the strength of the Higgs self-coupling and is analyzed in the following.

\subsubsection*{Enhanced Higgs self-coupling}
Interestingly, the scalar potential in the Einstein frame \eqref{potential} indicates that the Higgs cubic and quartic self-coupling is modified by gravitation effects. Around the electroweak vacuum and using \eqref{lambda} we find that
\be \lambda_{\rm eff}(m_{\rm EW}) = 0.13+\xi^2(m_{\rm EW})\alpha(m_{\rm EW})^{-1}.\ee
It is a distinctive deviation from the SM prediction. In particular, the Higgs quartic self-interaction at the EW scale is controlled by the parameter
\ba \lambda_4(m_{\rm EW}) &=& \lambda_{\rm eff}(m_{\rm EW})\cos^4\theta(m_{\rm EW})\cr
&\approx& (0.13+\xi^2 \alpha^{-1})(1-12\mpl^{-2}\xi^2 v^2)\big|_{\rm EW} .\ \ea

A sharp {\it prediction} of this mode is that the  Higgs self-coupling is greater than the SM prediction. Presently, there is no direct measurement of the Higgs self-coupling. The next generation of particle colliders are commissioned to directly measure the Higgs self-coupling in multi-Higgs processes. It is interesting to note that perturbativity ({\it i.e.} $\lambda_4\lesssim 1$) imposes a strong upper bound on the value of the Higgs non-minimal coupling parameter. For instance, if the Weyl field is responsible for cosmic inflation, then $\alpha\sim 10^8$ and we find an upper bound as $|\xi_{\rm EW}|\lesssim 10^4$. This is the tightest bound on the non-minimal coupling parameter in the literature. In general, as we could see in this framework with the SM interacting with gravity, collider experiments constrain the parameter space in the gravitational sector and favor regions where $\xi^2\alpha^{-1}\lesssim {\cal O}(1)$.

\subsubsection*{Gravitationally uplifted flat directions}
Another distinctive prediction in this framework is that, with the gravitational effects and a Higgs-like mechanism,  all scalar fields receive non-zero masses. Moreover, depending on the sign of the non-minimal coupling parameters, they might also develop a non-zero VEV. Different scenarios are studied in this part. 

We consider a generic scalar field $\psi$ with (essential) non-minimal coupling parameter $\xi_\psi$. It is interesting to note that, assuming no other potential term, then there will be a non-trivial potential  for $\psi$ in the Einstein frame
\be V_E\supset e^{-2\tilde\chi}\Big[{\textstyle\frac{1}{4}}\xi_\psi^2\alpha^{-1}\psi^4 -{\textstyle\frac{1}{2}}\xi_\psi\alpha^{-1}\big(\mpl^2 e^{\tilde\chi}-m^2\big)\psi^2\Big].\ee
Thus, the gravitational effects induce an effective self-coupling and mass parameter. 

We observed previously that the Wely field receives a non-zero VEV as the Higgs field develops a VEV \eqref{chi-vev}. Then, the field $\psi$ receives an effective mass parameter 
\be \mu^2_\psi = \mu^2_\phi\frac{\xi_\phi-\alpha\frac{-\mu^2}{m^2}}{\lambda+\xi_\phi\frac{-\mu^2}{m^2}}\xi_\psi\alpha^{-1}\approx  \mu^2_\phi\frac{\xi_\phi\xi_\psi}{\alpha\lambda}.\ee
We recall that $\mu_\phi^2<0$ and $\alpha,\lambda>0$. Depending on the sign of $\xi_\phi\xi_\psi$ two scenarios can be imagined. 
\paragraph*{i.  $\xi_\phi\xi_\psi<0$}: The field $\psi$ is settled at zero field value. However, its flat direction is lifted and it receives a mass
\be m^2_\psi = \mu_\psi^2\approx -{\textstyle\frac{1}{2}} \alpha^{-1}\xi_\psi\xi_\phi (125 {\rm GeV})^2.\ee

\paragraph*{ii.  $\xi_\phi\xi_\psi>0$}: The field $\psi$ is tachyonic around the origin and its receives a non-zero VEV
\be  \psi_0^2 = -\mu_\psi^2\xi^{-2}_\psi\alpha  \approx \xi^{-1}_\psi\xi_\phi(246 {\rm GeV})^2.\ee
The mass around this minimum is
\be  m_\psi^2 = -2\mu_\psi^2\approx \xi_\psi\xi_\phi \alpha^{-1}(125 {\rm GeV})^2.\ee
The gravitations effects induce a slight non-zero VEV and consequently break any symmetry under with the scalar field is charges.

In either cases, we observed that gravitational couplings induce a non-zero mass for scalar fields and lift the flat directions in the Einstein frame (if there was any in the Jordan frame). Furthermore, depending on the sign of non-minimal parameters, they might also develop a non-zero VEV through gravitational effects and spontaneously break symmetries if they are charged. 
Essentially, the non-zero VEV of the Weyl field, which itself is induced by the non-zero VEV of the Higgs field, induces a non-zero VEV to any scalar field (non-minimally coupled to gravity).  As can be explicitly observed, the dominant Higgs-like mechanism in a sector sets the relevant scales in the other sectors.  The Weyl field plays the role of a portal among different sectors. The significant prediction is that there is no scalar field with flat direction possibly no associated symmetry is preserved in a theory of gravity. It is a genuine gravitational effect.
 
\subsection*{V. Conclusion}
In this note we proposed a general gravitational framework with scalars non-minimally coupled to a Ricci-squared theory of gravity as it is implied by renormalization theory. We found that there is a two-parameter family of models with rich dynamics. We were interested in region of the parameter space where the emergent Weyl field is the inflaton. We found that, at early times, the Higgs field in the Einstein frame had a large effective mass. It quickly relaxed to its small field values through damped oscillations prior to inflation. It alleviates the metastability problem of the Higgs potential and explains why the Higgs field could be trapped in the shallow electroweak vacuum. 

This framework has two predictions. Firstly, we found that the Higgs self-coupling receives contribution from the gravitational sector and thus is enhanced. Any deviation in future measurements can be naturally explained through omnipresent gravitational couplings with no need of new physics. On the other hand, perturbation theory in low scale physics constrains the gravitational parameter space. 

Secondly, we observed that natural gravitational couplings lift flat directions along moduli fields. Moreover, in some part of the parameter space, scalar fields develop non-zero VEVs through pure gravitational effects. If they are charged, they gravitationally spontaneously break symmetries. They Weyl field plays the role of a portal that carries the effect of the dominant Higgs-like mechanism in one sector to the other sectors. If a non-minimally coupled scalar receives a non-zero VEV, all other scalars with non-minimal couplings receive non-zero masses and possibly non-zero VEV's.  All the above observations  are universal and generic as are induced by ubiquitous and natural extensions in the gravity sector. 

\paragraph*{Acknowledgments}
This work is supported by the research deputy of SUT.

\paragraph*{Note added:} This article is an expanded and a published version of \cite{Torabian:2014nva} initially presented in IPM $1^{st}$ Topical Workshop on Theoretical Physics.

\ \\$^*$ Electronic address: mahdi.torabian@sharif.ir

\end{document}